\newcommand{\beq}{\begin{quote}}
\newcommand{\enq}{\end{quote}}
\newcommand{\be}{\begin{equation}}
\newcommand{\en}{\end{equation}}

\documentclass[11pt]{amsart}
\usepackage{geometry}                
\usepackage{graphicx}
\usepackage{amssymb}
\usepackage{epstopdf}
\title{Robert Gibson and  teaching  philosophy of nature  in  18th century Ireland}
\author{Michael Nauenberg\\
Department of Physics\\
University of California, Santa Cruz, CA 95064} 

\begin{document}
\maketitle
\begin{abstract}
The dissemination of natural philosophy in the 18th-century, which was based primarily on Newton's pioneering work in mechanics, optics and
astrophysics,  is presented as seen through a remarkable textbook written by a little known
Irish mathematics teacher,  Robert Gibson.  Later, he became  the  deputy surveyor general of Ireland (1752-1760).

\end{abstract}
\section* {Introduction}

       Some time ago,  while  browsing  in an antique book shop in London, I came across 
a book with the interesting title {\it A Course of Experimental Philosophy;  Being an Introduction to the true
philosophy  of Sir Isaac Newton}.  Intrigued,  I looked for the author's name, and found that  he
was listed as Robert Gibson,  teacher of Mathematics, and that his book was printed 
in  Dublin, Ireland, in 1755.  Most remarkable,  the frontispiece  featured a drawing of an air pump (see Fig.1),   which puzzled me, because
I was aware that Newton had not even  mentioned  experiments with such a pump in his {\it Principia}.  Later,  I found that
this  particular  pump was  built by Francis  Hauksbee,  who became a curator of experiments at the London Royal Society when Newton became president, after Robert Hooke's death  in 1703.

  \begin{figure}[htbp] 
   \centering
   \includegraphics[width=6in]{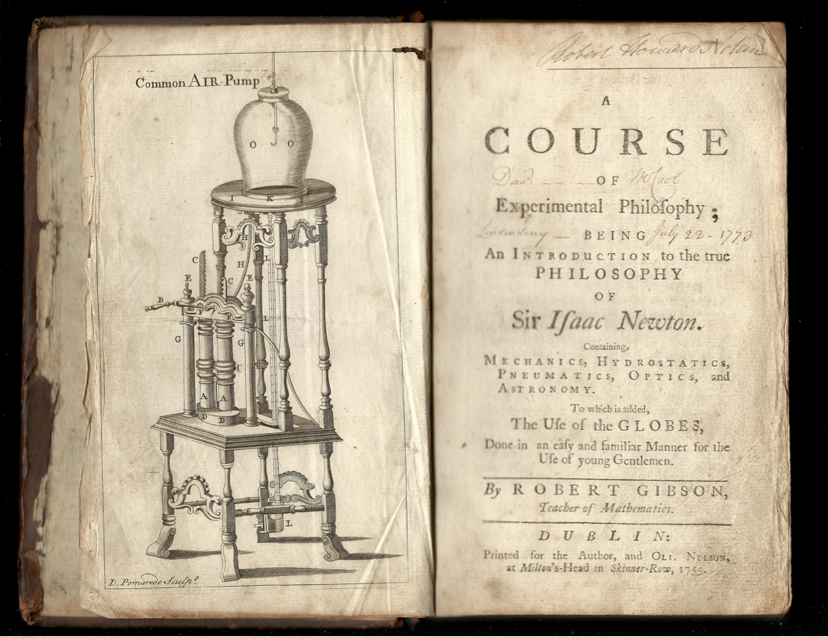} 
   \caption{ The opening two  pages  of Robert Gibson's book with a drawing of  Hauksbee's air pump.
   }
\end{figure}
In his preface, Gibson stated  the purpose of his book:

\begin{quote}
The Design then is, to explain in the most easy and concise Manner, so much of the Science, as may enable young Gentlemen,
or Men of Business, to form a general Idea  of the Elements, or the Rudiments of it:  Experience in my Profession has convinced
me  that few desire more,  or will give themselves the Trouble  to form a critical Notion of every Part of it;  because they must 
content with voluminous Tracts, which contain many abstruse mathematical Reasonings, which require  a previous mathematical
Knowledge of the Elements of Euclid, Conic-Sections, Algebra and Fluxions;
and Indeed a general and concise Description  or Account of any Art or Science, is best adapted to answer the Views and Ends of the
Greatest  part of Readers.

As for the Execution of the Work, the subject is for the most Part illustrated by Experiments that carry with them Evidence, sufficient to
satisfy the most curious Mind: There are some  Geometric Demonstrations which if the reader would understand  will require the 
Assistance of Euclid; but as these are  only a few they may be passed by,  by such as are  ignorant of Geometry, taking the Premises
of the Propositions for granted . . .
\end{quote}

 I also  examined the content of several  other $18th$ century books  that were written to explain 
Newton's  concepts,  and I found that most of them were based on an approach similar to Gibson's:  to  discuss  the experiments 
that led to Newton's fundamental concepts in mechanics and  optics,  while leaving out most of the mathematical  arguments 
that supported them. Moreover,  as with present textbooks,  the authors copied from each other,  which accounts for  the
great deal of similarity in their presentations.  In his introduction, Gibson admitted that,
\beq
In drawing up this course, I have not scruple to take whatever I judge might best answer my Purposes from the best authors.
\enq
Hence, the contents and the presentation in his  book  is also characteristic of many  other
books  that familiarized  people in the $18th$  century  with   Newton's  achievements.

After  Gibson's   preface, there are six pages
listing {\it subscribers names}  and their professions.  These were people from all walks of contemporary life, who bought a copy before the book's printing (see Fig.2).

  \begin{figure}[htbp] 
   \centering
   \includegraphics[width=6in]{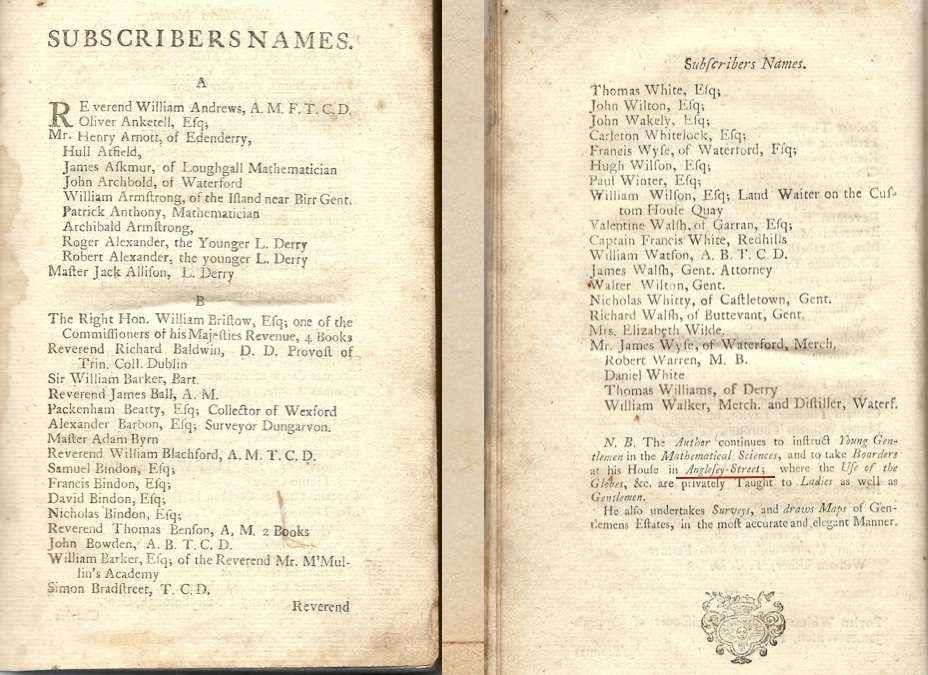} 
   \caption{ A partial list of the subscribers to Gibson's book
   }
\end{figure}
Similar lists appear also in other textbooks written at this period.  At the end of his list, Gibson included  an announcement  
that 

\beq
The {\it Author} continues to instruct Young
Gentlemen in the {\it Mathematical Sciences}, and  takes  {\it Boarders} at this House in {\it Anglesey-Street} 
\footnote{
{\it Anglesea St.},  as it is now called, is not far from  St. Stephen's Green, in an area  called Temple Bar near the  river Liffey.  
I went there to look  for some plaque that would commemorate the location
 where Gibson had given  his lessons,  but I found only plenty of  bars.},
 where the Use of the Globes,
$\&c.$ are privately Taught to {\it Ladies} as well as {\it Gentleman}.
\enq

Later, Gibson became  the deputy  surveyor general of Ireland (1752-1760), and  was well know as the author of {\it A Treatise of Practical Surveying} that was published in several editions, including two in the American colonies, one in Baltimore and another one in Philadelphia
\cite{survey}.
I think it is likely that George Washington, who was a surveyor at the time,  learned his trade by reading Gibson's textbook. Unfortunately
Washington's library has been lost, so I could not verify my conjecture.

\section*{Teaching Newton's Philosophy of Nature in Europe}

Two of the  main figures that taught and promulgated  Newtonian principles in Europe were   John Theophilus Desaguliers,   and Willem
Jacob 'sGravesande.  As Huguenots, Desaguliers family fled to England when he was 11 years old. In 1712  he succeed John Keill in reading lectures on experimental philosophy at   Oxford, and  later he followed Hauksbee to become curator of experiments at the Royal Society in London. s'Gravesande  was Dutch, and in 1715 he  spent a year in England where he attended  demonstrations of experiments
by Desaguliers. In 1717 he became professor of Physics and Astronomy in Leiden, where he began to introduce
the Newtonian experimental philosophy in the Netherlands.  
A third important figure in the early popular dissemination of  Newton's work  was  Voltaire,  who in 1726 self-exiled went
to England  to avoid being sent to the Bastille for his sarcastic  writings that offended the French authorities.
In his famous {\it  Philosophical Letters}  he wrote,  in reference to the dispute between the Cartesian and Newtonian philosophy
of nature,  that : 

\beq
``In Paris they see the universe as composed
of vortices of subtle matter, in London they
see nothing of the kind.
For your Cartesians  everything is moved
by an impulsion you don't really understand,
for Mr. Newton  it is by gravitation, the cause
of which is hardly better known.  In Paris you
see the earth shaped like a melon, in  London
it is flattened on two sides"
\enq
The  comment about  the shape of the earth was made regarding  the controversy between Giovanni Domenico 
Cassini's measurements  of the length  of a degree of the meridian,   indicating that the
earth was elongated at the poles, and Newton's theoretical calculation in
 his {\it Principia}, indicating  the  opposite.  An expedition,  led by Pierre Louis Maupertuis   to perform such measurements
 near the artic circle in Lapland,   confirmed Newton's prediction that the earth was an oblate spheroid, flattened at the poles. 
 
After returning to France, Voltaire published {\it The elements of Sir Isaac Newton's philosphy} under his sole name, but in the frontispiece
 he revealed his muse  and unnamed co-author:  \'{E}milie du Chatelet  (see Fig.3).  She was sufficiently  versed
 in  mathematics  to understand the {\it Principia},  and translate it  into  French.  To Frederick II
 of Prussia, Voltaire admitted that
 \beq
 Minerva dictait et j'ecrivais \footnote{ Minerva (\'{E}milie) dictates, and I write.}.
 \enq

  \begin{figure}[htbp] 
   \centering
   \includegraphics[width=4in]{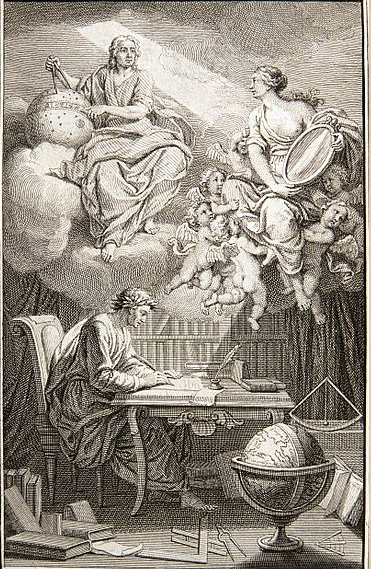} 
   \caption{ Frontispiece of Voltaire's {\it The Elements of Newton's Philosophy}.
Voltaire's manuscript is illuminated by seemingly divine light coming from Newton himself, reflected down to Voltaire by a muse, representing  \'{E}milie du Chatele.
   }
\end{figure}
Desaguleirs wrote a very influential textbook, {\it A Course of Experimental Philosophy}, that was translated into several languages. 
 In his introduction he explained  with an amusing tale, the need to understand the fundamental laws of nature,
 
 \beq
 About two years ago a man proposed an engine to raise by one man's work about ten times more water than was possible for 
 a certain height in a certain time; for which he wanted an Act of Parliament,  and got a report of the Committee, appointed to examine
 the matter that he had made out the allegation of his petition.  If this had passed,  a great many persons were ready to subscribe
 considerable sums of money to the project; which of course would all have been lost, and perhaps some families ruined,
 but a nobleman, who understand the nature of engines very well, knowing the impossibility of what was proposed, threw out
 the Bill.
 Our legislator may make laws to govern us, repeal same, and enact others, and we must obey them;  but they cannot alter
 the laws of nature; nor add or take away one {\it iota} from the gravity of bodies.
 \enq
 
  In the preface of his  second volume he wrote:

\beq
This second volume has more need of an apology than a preface, upon many accounts; first account of the time
that it has been delayed, when it should have immediately follow the first book to decide the question, which has been
a subject of dispute about 59 years; the gentleman of Germany, Italy and Holland measuring  Force by the product
of the mass into the square of the velocity of the body $[mv^2]$,  and those of France and England measuring Force by the
product of the mass into the simple velocity $[mv]$.
\enq
Of course, at the end Desaguliers  should have written  that the measure of force is the product of the mass times the {\it change} of velocity, when
this force is an instantaneous {\it impulse}, but confusion about this subject remains up to the present time
\footnote{In definition  6  of his {\it Principia}, Newton  defined the {\it motive} force in the familiar form as the product of mass
times  the acceleration,  and in Proposition 6 he gave  a precise  mathematical expression for this force.} \cite{michael}.  
Then he continued:

\beq
I could not quit my conviction in favour of the old opinion, as it was supported by demonstration; but yet could not find
any want of accuracy in several of the experiment I examined, which were made to prove the new opinion; neither could
I find any fallacy in the reasoning from those experiments; tho' I thought it must be want of penetration in me that I could not
perceive it, supposing the both opinions could not be true.
\enq
 For example,  \'{E}milie du Chatelet    did experiments dropping weights from different
heights,   and found that the impact on the ground  was proportional to the height $h$, which, according to Galileo,
was proportional to
$v^2 $. 
Finally, the resolution of this controversy  came  with the realization that the experiments  that supported   the notion  that  
force equals $mv^2$  were actually measuring the change in energy rather than in momentum.  Desaguliers concluded that
\beq
. . .  examining everything again with care, found that the philosophers on both sides were right in the main;  but only
so far wrong as they attributed  to their adversaries opinions, which they had not: and that the whole was only a dispute about
words; the contending parties meaning different things by the word  force
\enq

In the preface of the first edition of  his book, {\it Mathematical elements of Natural philosophy confirmed by experiments, or an introduction
to Sir Isaac Newton's philosophy}, 's Gravesande  wrote:

\beq
In order to render the study of natural philosophy as easy and agreeable as possible,  I have thought to illustrate everything 
by experiments, and to set the very mathematical conclusions before the readers eyes by this method.
\enq

He describe  its contents  as follows;
\begin{quote}
The whole work is divided into four books. The first treats of body
in general, and the motion of solid bodies. The second of fluids.  What belongs
ot light is handled in the third.  The fourth explains the motion of celestial
bodies and what has a relation to them on earth.
\end{quote}
                                                          
His book was  illustrated with realistic  drawings  of his experimental equipment  (see Fig.4),  which, by his own account,  was built by
John van Musschenbroek, whose son, Pieter, succeed 'sGravesande as a professor in Leiden in 1739.

\begin{figure}[htbp]
\centering
\includegraphics[width=10cm]{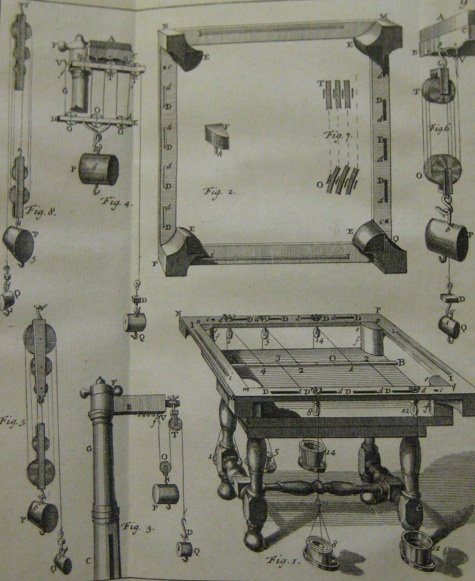}
\caption{ A detailed drawing of the apparatus of 'sGravesande for demonstrations in mechanics.
}
\end{figure}
\begin{figure}[htbp]
\centering
\includegraphics[width=8cm]{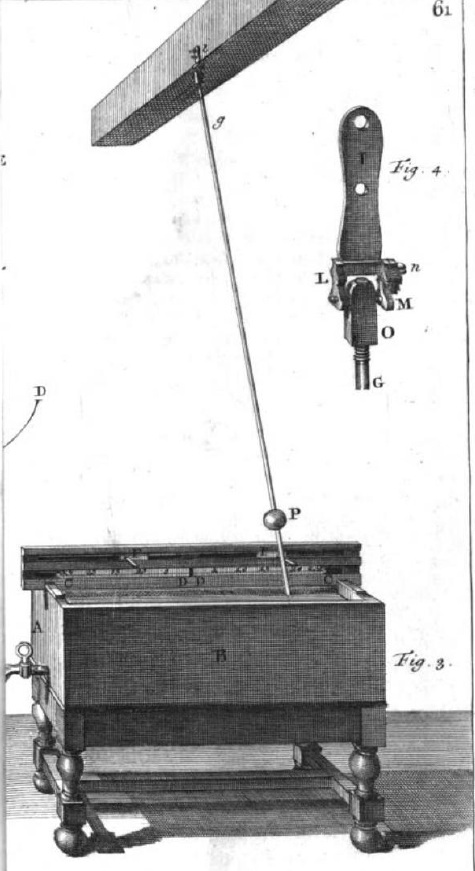}
\caption{ Apparatus to demonstrate the damping of a pendulum immersed in a fluid
}
\end{figure}

'sGravesande's book, originally written in Latin, was very influential,  went through several editions, and it was translated into several languages, e.g. its  second edition 
was translated into English by Desaguliers, and  Voltaire travelled  to Leiden to learn 
directly from 'sGravesande's  lectures on Newton's physics.
In Ireland,  Richard Helsham, who became president of the Royal college of physicians in 1716,   gave lectures on experimental
philosophy at the university of Dublin from 1724 to 1740. Hew also wrote a textbook on this subject, entitled
{\it A course on lectures in Natural Philosophy}, published in1739, which was translated into Latin by a Jesuit, and printed in 
Vienna in 1769 under the title {\it Physica experimentalis Newtoniana}.  Helshman's   comment that
\begin{quote}
``The purpose of science education is to appreciate examples of God's workmanship in the natural world",
\end{quote}
illustrates  another important  motivation for the widespread dissemination of  Newton's work.

In England Henry Pemberton, who supervised the third edition of the {\it Principia},  published  his {\it View of Sir Isaac Newton's 
philosophy} in 1728, and John Keill, whose vigorous defense of Newton against Leibniz during the priority dispute earned him the name of
``Newton's pit bull",  published an {\it Introduction to Natural Philosophy: or philosophical lectures}.  In contrast, in  Italy, where Galileo had set in motion the wheels  that ultimately led
to Newton's  development  of mechanics, there appeared a book entitled  {\it Newtonianismo per le dame},
by  a light-weight scholar,  Francesco Algarotti, 
who dedicated it  to Fontanelle, whose style  of presentation - the philosopher
instructing a lady of high rank during conversations with her in the evenings
- he appropriated from Fontenelle.\footnote{
In 1686,  a year before the apperance of Newton's {\it Principia},  Bernard le Bouvier de Fontenelle  published an influential  book, {\it Entretiens sur La Pluralite des Mondes}, written  in the style  adopted by Algarotti.  In his book, Fontenelle presented good arguments
for the existence of planets around stars other than the sun, but only very recently (since 1996) 
have such exo-planets  been observed. 
}  As an example, consider her response to Algarotti's 
 description  of  Newton's inverse square law of gravity:

\begin{quote}
I cannot help thinking, said the Marchioness, that this Proportion in the Squares of the Distances of Places, or rather of Times, is observed even in Love. Thus after eight Days Absence, Love becomes sixty four Times less than it was the first Day, and according to this Proportion it must soon be entirely obliterated: I fancy there will be found, especially in the present Age, very few Experiments to the contrary.
\end{quote}
Not surprisingly,  an anti-Enlightenment poet, Giuseppe Parini, in a satirical poem {\it Il giorno}, portrayed a lady listening ecstatically to her cicisbeo\footnote{ In Italy a {\it cicisbeo} or {\it cavalier servente}, was the lover of a married woman, who attended her at 
public entertainment,  and other occasions.} as he expounds upon calculus, mass and inverse ratio at the dining table. For this poet, the transformation of infinitesimal calculus and universal gravitation into a topic of amorous conversations was a sign of the moral corruption of Italian aristocratic society. 
 
For further details on the dissemination of Newton' philosophy of nature in the 18-th century,  see references \cite{moti} and \cite{pulte}.
It is also  interesting to contrast the  presentation of natural philosophy in that century,  with the 
corresponding teaching in the first half of the  previous one \cite{reif}.

\begin{figure}[htbp]
\centering
\includegraphics[width=10cm]{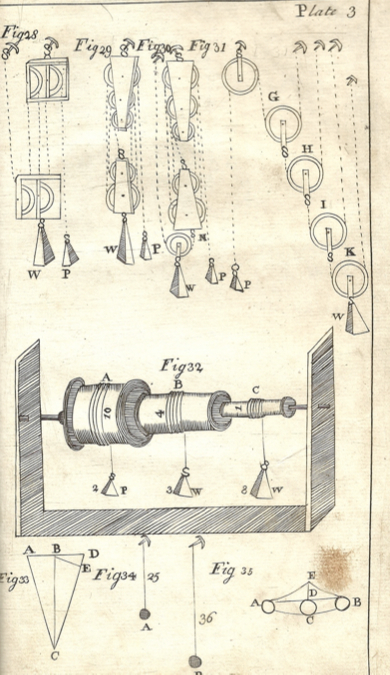}
\caption{Illustrations of the mechanical advantage due to pulleys.
}
\end{figure}

\section*{ Some illustrations in Robert Gibson's textbook.}

In accordance with  most 18-th century textbooks on Newton's experimental philosophy,  the main topics
that Gibson  covered  in his  book are  mechanics, fluids, optics,
and observational astronomy. These topics are illustrated by the figures printed at the end
of his book,  many of which were  copied from other books.  For example, the reproduction shown in Fig.6, illustrating 
 the mechanical advantage of 
compound pulleys,  is similar to the corresponding figures in  Hauskbee's book. In constrast,  Fig.7 illustrates Gibson own
rough hand drawings of fluid jets  and the mechanics
of water pumps.  The studies of these jets can be traced  back  to Leonardo da Vinci at the end of the
15-th century (see Fig.8).  In particular, Gibson's  Fig.4.5  contains a diagram, based on Descartes'  {\it Geometrie} - the geometrical
evaluation of the square root of the product of two quantities -  that also appeared  in all the books that discuss this problem. It gives 
the horizontal distance $x$ that a jet travels,   given the height $h$ of the water above the opening, and the height $y$ of this opening
above ground.  The result, $x=2\sqrt{hy}$, is based on a principle  first enunciated in 1644 
by Evangelista Torricelli \footnote{ For example, for the opening at $B$, $FB=\sqrt{GB\times y}$, where $y$ is the height of the
opening above the ground, and ccording to Toricelli,  the distance $x$ of the emerging  jet  is $x=2 FB$. }.
Surprisingly,  Newton did not get this result correctly  until the  second edition of his  {\it Principia} in 1713. 

\begin{figure}[htbp]
\centering
\includegraphics[width=10cm]{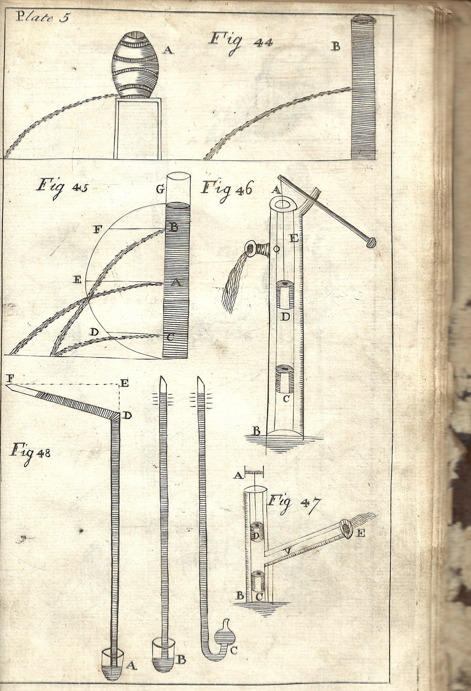}
\caption{Gibson's illustrations of water jets, Figs. 44 and 45, and the
operation of water puimps, Figs. 46 and 47.
}
\end{figure}

\begin{figure}[htbp]
\centering
\includegraphics[width=10cm]{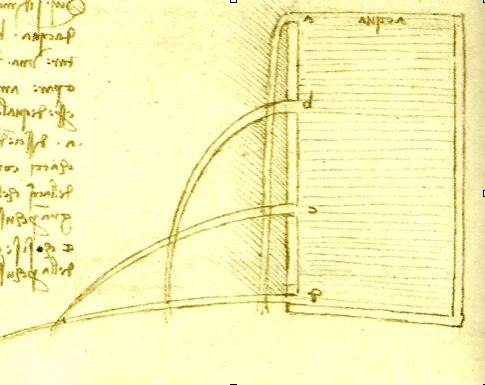}
\caption{ Leonardo da Vinci drawing of one of his pioneering  experiment in the 15-th
century,  with water jets emerging from holes  on the side of  a tank.
}
\end{figure}

Under the heading, {\it The flux of water from reservoirs through orifices and pipes}, Gibson ennunciated succintly Torricelli's principle:
\begin{quote}
If water flows through an orifice in the bottom of a vessel which is kept constantly full, or to the same height; the velocity
with which it flows out, is as the square root of its height above the orifice.
\end{quote}

He then proceeded  to describe an experiment to verify this principle:
\begin{quote}
Let there be two vessels alike in all things, except that one is four times as tall as the other, or let the height of  A be 20
and of B be 5, each having a circular  orifice  of $1/3$ part of an inch in the bottom.  If these vessels be filled with water and set running,
the water being constantly supplied  above as fast as it runs out below;  the taller vessel will discharge 21  ounces in a quarter
of a  minute, and the latter 11 ounces: Therefore, the velocity with which the water  flows out of the taller vessel is to the velocity 
wherewith it flows out of the shorter one, as 21 to 11, that is as 2 to 1 nearly; which numbers are the square roots of  4 and 1
which expresses the proportion of the heights of the water above the orifices
\end{quote}

A few pages later,  he discussed a corresponding experiment with a pipe inserted horizontally in a hole of the vessel.
\begin{quote}
The velocity wherewith water flows out of a cylindrical pipe inserted horizontallly in the side of a vessel, is as the square root of the height 
of the water, above the place of the pipe's insertion directly, and the square root of the lenght of the pipe inversely. For the place of the
orifice may be looked on as if it were in the bottom of the vessel, since no water under it can flow out;  and the same velocity where with the 
water flows out  of the cylindrical pipe,  with the very same velocity it flows in it a the other end,
that is, it will be as the square root of the height above the orifice:  but the water in the pipe becomes a Clog, and impedes the
velocity rushing in  at  the pipe, and the longer this pipe is the greater  the impediment will grow, and of course the less will be the 
velocity of the water in the pipe, and this  is {\it  found to be inversely as the square root of the pipe's length} . . .[the italics are mine]
\end{quote}
Subsequently,   Gibson  presented  an experiment supposedly  verifying  the hypothesis that the velocity of the water in the pipe varies inversely
as the square root of the length of the pipe.  His account, however,  is identical to an incorrect experiment described
 in the section on hydrostaticks in  Helshman's 
textbook.  Actually, the velocity in the pipe varies inversely as  its length. This  dependence is   due to the viscosity
of fluids, and  not to the inertial effects of a {\it clog}. It was first  established correctly  more
than a century later by the experiments of  a German engineer,   Gotthilf Hagen, and  a French physician Jean Poiseuillie\cite{dariol},\cite{hagen}.

 It is interesting to compare Gibson figures for a fluid jet,  with the corresponding ones in Desaguliers book, which shows
a realistic drawing of a device for the actual experiment (see Fig.9).  This figure  shows that  jets emerging vertically 
from holes of different radius of aperture reach different heights, which is correct (as I have observed in such an experiment), 
but disagrees with  Torricelli's principle.  Desaguliers figure provides direct evidence that  he  verified at least some of the 
experiments that are discussed
in his book, while Gibson, like many other textbook writers (then as well as up to the present time)
  was content with  reporting  the results  published by other authors..
  
  \begin{figure}[htbp]
\centering
\includegraphics[width=10cm]{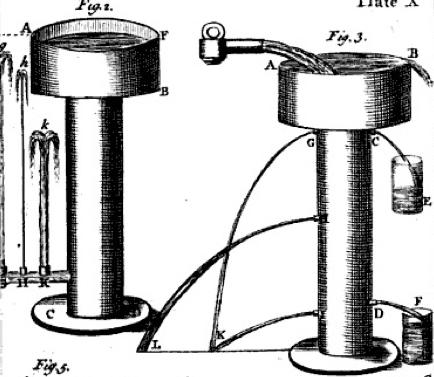}
\caption{ Illustrations of jets in Desaguliers book
}
\end{figure}

  \begin{figure}[htbp]
\centering
\includegraphics[width=10cm]{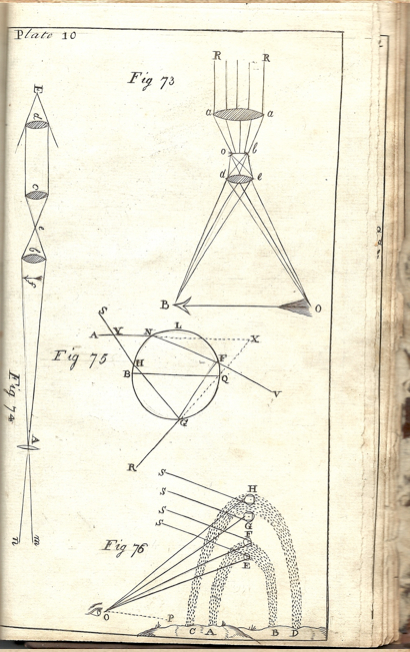}
\caption{  Illustrations for some  optical ray trajectories in lenses, Figs.73 and 74, and
in the formation of the rainbow, Figs. 75 and 76.
}
\end{figure}
In Fig.10, Gibson 
illustrated various optical phenomena, some of which were copied  directly from Newton's {\it Opticks}. For example, in his 
Figs. 75 and 76, the diagrams show graphically  the path of sun rays that form  the rainbow, which appear
 when sunlight is refracted and reflected  by water drops. Gibson did not even bother
 to change the labels in Newton's  corresponding diagrams that  appear in Book 1, Part 2, Prop. 9, as  Figs. 14 and 15.  Ironically, the original calculations for the rainbow were first carried out  by Descartes, who
  published his graphical method to obtain the scattering angle of the sunlight in
a  supplement, entitled {\it Meteors}, of his  famous treatise {\it ``Discourse on the Method of rightly conducting one's reason and of seeking truth in  the Sciences"}, where one finds similar figures.
The other two figures illustrate the formation and magnification of images by lenses,  a subject  of great practical importance  for the construction
of optical instrumentsm,  and undoubtedly much appreciated by lens makers.

The last quarter of  Gibson's book gives an excellent but mainly qualitative  introduction  of planetary astrophysics,  based on Book 3 of Newton's {\it Principia} .
\clearpage

\section*{ Acknowledgments}
I would like to thank  Enda McGlynn and Kelly Patrick for  information about Robert Gibson and  Richard
Helshman.

\end{document}